\documentclass{nature1} 
\usepackage{graphicx}
\title{The type Ia supernova SNLS-03D3bb from a super-Chandrasekhar-mass white dwarf star}
\author{D.~Andrew~Howell$^1$, Mark~Sullivan$^1$, Peter~E.~Nugent$^2$,
  Richard~S. Ellis$^3$, Alexander~J.~Conley$^1$, Damien~Le~Borgne$^4$,
  Raymond~G.~Carlberg$^1$, Julien~Guy$^5$, David~Balam$^6$,
  Stephane~Basa$^7$, Dominique~Fouchez$^8$, Isobel~M.~Hook$^9$,
  Eric~Y.~Hsiao$^6$, James~D.~Neill$^6$, Reynald~Pain$^5$,
  Kathryn~M.~Perrett$^1$, Christopher~J.~Pritchet$^6$ }
                                                          
\usepackage[super,sort]{natbib} 

\begin{document}

\maketitle
\begin{affiliations}
\item Department of Astronomy and Astrophysics, University of
Toronto, 60 St. George Street, Toronto, ON M5S 3H8, Canada
\item Lawrence Berkeley National Laboratory, Mail Stop 50-232, 1
Cyclotron Road, Berkeley CA 94720 USA
\item California Institute of Technology, E. California Blvd.,
  Pasadena, CA 91125, USA
\item DAPNIA/Service d'Astrophysique, CEA/Saclay,
91191 Gif-sur-Yvette Cedex, France
\item LPNHE, CNRS-IN2P3 and University of
Paris VI \& VII, 75005 Paris, France
\item Department of Physics and Astronomy, University of
Victoria, PO Box 3055, Victoria, BC V8W 3P6, Canada
\item LAM CNRS, BP8, Traverse du Siphon, 13376 Marseille Cedex 12, France
\item CPPM, CNRS-IN2P3 and University Aix Marseille II, Case 907, 13288 Marseille Cedex 9, France
\item University of Oxford Astrophysics, Denys Wilkinson Building, Keble Road, Oxford OX1 3RH, UK

\end{affiliations}

\newcommand\ion[2]{#1$\;${\small\rmfamily\@Roman{#2}}\relax}%
\newcommand\nodata{ ~$\cdots$~ }%
\newcommand\arcdeg{\mbox{$^\circ$}}%
\newcommand\degr{\arcdeg}%
\newcommand{\etal}{et~al.\,}
\newcommand{\Msun}{${\rm M}_\odot$}          
\newcommand{\drp}{\Delta m_{15}(B)} 
\newcommand{\Ni}{$^{56}\rm Ni$} 
\newcommand{\Co}{$^{56}\rm Co$} 
\newcommand{\Fe}{$^{56}\rm Fe$} 
\newcommand{\kps}{km s$^{-1}$}
\newcommand{\kms}{km s$^{-1}$}
\newcommand{\MCh}{$M_{\rm Ch}$}
\newcommand{\hof}{H\"oflich}
\newcommand{\eg}{e.\,g.\,}
\newcommand{\ie}{i.\,e.\,}
\newcommand{\OI}{\ion{O}{1}}
\newcommand{\SiII}{\ion{Si}{2}}
\newcommand{\Simain}{\ion{Si}{2} 6150\AA}
\newcommand{\Sifour}{\ion{Si}{2} 4000\AA}
\newcommand{\SII}{\ion{S}{2}}
\newcommand{\CaII}{\ion{Ca}{2}}
\newcommand{\Camain}{\ion{Ca}{2} 3800\AA}
\newcommand{\Ca}{\ion{Ca}{2}}
\newcommand{\TiII}{\ion{Ti}{2}}
\newcommand{\MgII}{\ion{Mg}{2}}
\newcommand{\iab}{i\arcmin (AB)}
\newcommand{\us}{$u$*}
\newcommand{\gp}{$g$\arcmin}
\newcommand{\rp}{$r$\arcmin}
\newcommand{\ip}{$i$\arcmin}
\newcommand{\zp}{$z$\arcmin}
\newcommand\arcmin{\mbox{$^\prime$}}%
\newcommand\arcsec{\mbox{$^{\prime\prime}$}}%
\newcommand\araa{Ann. Rev. Astron. Astrophys.}%
\newcommand\apj{Astrophys. J.}%
\newcommand\aj{Astron. J.}%
\newcommand\apjl{Astrophys. J. Lett.}%
\newcommand\apjs{Astrophys. J. Suppl.}%
\newcommand\aap{Astron. \& Astrophys.}%
\newcommand\mnras{MNRAS}%
\newcommand\nat{Nature}%
\newcommand\iaucirc{IAU~Circ.}%
\newcommand\gca{Geochim.~Cosmochim.~Acta}%
\newcommand\aaps{Astron. \& Astrophys. Suppl. Ser.}%
\newcommand\pasp{PASP}

\begin{abstract}  
  The acceleration of the expansion of the universe, and the need for
  Dark Energy, were inferred from the observations of Type Ia
  supernovae (SNe Ia)\citep{1998AJ....116.1009R,1999ApJ...517..565P}.
  There is consensus that SNe Ia are thermonuclear explosions that
  destroy carbon-oxygen white dwarf stars that accrete matter from a
  companion star\citep{1960ApJ...132..565H}, although the nature of
  this companion remains uncertain.  SNe Ia are thought to be reliable
  distance indicators because they have a standard amount of fuel and
  a uniform trigger --- they are predicted to explode when the mass of
  the white dwarf nears the Chandrasekhar
  mass\citep{1931ApJ....74...81C} --- 1.4 solar masses.  Here we show
  that the high redshift supernova SNLS-03D3bb has an exceptionally
  high luminosity and low kinetic energy
  that both imply a {\em super}-Chandrasekhar mass progenitor. 
  Super-Chandrasekhar mass SNe Ia should preferentially occur in a
  young stellar population, so this may provide an explanation for the
  observed trend that overluminous SNe Ia only occur in young
  environments\citep{1996AJ....112.2391H,2006sullivan}.  Since this
  supernova does not obey the relations that allow them to be
  calibrated as standard candles, and since no counterparts have
  been found at low redshift, future cosmology studies will have to
  consider contamination from such events.

\end{abstract}

\begin{figure*}
\includegraphics[width=6in]{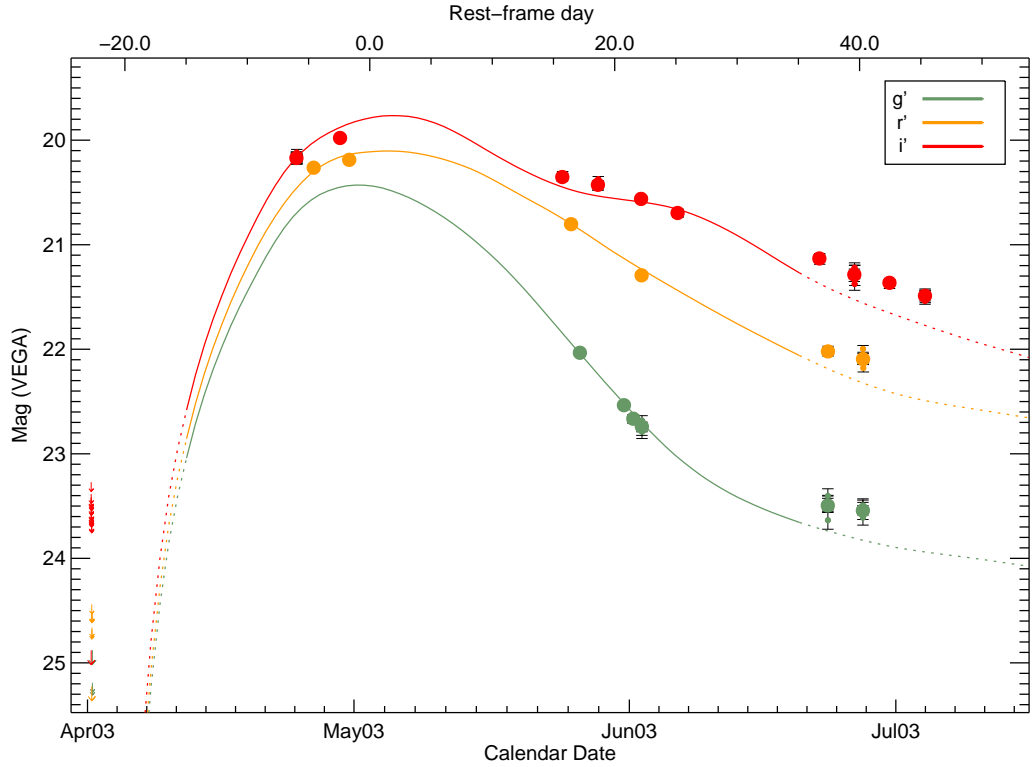}
\caption{The lightcurve of SNLS-03D3bb.  
  We fit k-corrected \citep{2002PASP..114..803N} template  lightcurves
  to the observed photometry of  SNLS-03D3bb, then transform the  peak
  magnitudes back to the Johnson-Cousins   \citep{1990PASP..102.1181B}
  $BV$  magnitudes in  the Vega  system.  We  find  peak magnitudes of
  $B=20.35$  mag and $V=20.50\pm0.06$ mag  from  a simultaneous fit to
  \gp\  and \rp\ data.  The error (s.d.) consists of 0.04 statistical error
  and 0.04 k-correction error.  A  lightcurve  template was fit  using
  the  stretch method\citep{1999ApJ...517..565P}  (stretching the time
  axis of a template lightcurve  by a stretch  factor, $s=1.13$).  The
  epoch of maximum  light  relative to  the  rest frame $B$   band was
  determined from a simultaneous fit to all of  the data.  At maximum,
  we only use the $V$ band value to compare to  other SNe, since it is
  the best constrained.  Data past +35 days  was not used in the  fit. 
  The arrows are three sigma upper limits.\label{lc}}
\end{figure*}

SNLS-03D3bb (SN~2003fg) was discovered on April 24, 2003 (UT) as part
of the Supernova Legacy Survey (SNLS).  Its redshift is $z=0.2440 \pm
0.0003$, determined from host galaxy [OII], [OIII], H$\alpha$, and
H$\beta$ emission lines.  A finding chart and observational details
can be found in the supplementary information (SI).  From the lightcurve
(Figure~\ref{lc}) we measure a peak magnitude in the rest frame
$V$ band, $V=20.50 \pm 0.06$ mag.  This corresponds to an absolute
magnitude of $M_V=-19.94 \pm 0.06$ ($H_0=70$ km~s$^{-1}$~Mpc$^{-1}$,
$\Omega_M=0.3$, flat universe).  SNLS-03D3bb falls completely
outside of the $M_V$ distribution of low-z SNe
Ia\citep{2006A&A...447...31A}, and is 0.87 mag (a factor of 2.2)
brighter than the median.  Note that neither changes in the Hubble constant nor
$\Omega_M$ significantly affect this brightness difference.  Asphericity may account for
variations in SN Ia luminosity at the 25\% level, but not a factor of
two\citep{2001ApJ...556..302H,2004ApJ...610..876K}.  SNLS-03D3bb also does not
follow the lightcurve width-luminosity
relationship\citep{1993ApJ...413L.105P} for SNe Ia that allows them to
be calibrated as standard candles --- it is too bright for its
lightcurve width (``stretch'', s=1.13) by $0.61\pm0.14$ mag ($4.4
\sigma$).

Type Ia supernovae are powered exclusively by the decay of \Ni\ and
its decay product \Co \citep{1969ApJ...157..623C}, requiring $\sim
0.6$ \Msun\ of \Ni\ to reproduce a normal SNe
Ia\citep{1984ApJ...286..644N, 1986ARA&A..24..205W,
  1992ApJ...392...35B,1995ApJ...443...89H}.  Since SNLS-03D3bb is 2.2
times overluminous, this implies that it has $\sim 1.3$ \Msun of \Ni .
Such a large \Ni\ mass is not possible if the progenitor is limited to
the Chandrasekhar mass (1.4 \Msun ).  Even models that burn the entire
1.4 \Msun\ to nuclear statistical equilibrium via a pure detonation
produce only 0.92 \Msun\ of \Ni , with the remainder comprising other
iron-peak elements\citep{1993A&A...270..223K}.  Since at least 40\% of
the SN Ia must be elements other than \Ni\ to reproduce observed
spectra\citep{1984ApJ...286..644N,1999MNRAS.304...67F}, this implies a
WD mass of $\sim 2.1$ \Msun .  Some authors find 
rapid rotation may support such a massive white 
dwarf\citep{2005A&A...435..967Y}.  The merger of two massive white 
dwarfs could also produce a super-Chandrasekhar 
product\citep{1994MNRAS.268..871T,2001ApJ...554L.193H}.

This simple estimation of the nickel mass is supported by a more
detailed calculation using the principle that the
luminosity at maximum light is proportional to the instantaneous rate
of radioactive decay\citep{1982ApJ...253..785A,1985Natur.314..337A}.  
The impied Ni mass is\citep{1995PhRvL..75..394N,1997PhDT.........7N}:
$M_{\rm Ni}=\frac{L_{\rm bol}}{\alpha \dot{S}(t_R)}$, where 
$L_{\rm bol}$ is the bolometric luminosity at maximum light (the
luminosity integrated from the ultraviolet to the infrared), and
$\alpha$ is the ratio of bolometric to radioactivity luminosities,
near unity. $\dot{S}$ is the radioactivity luminosity per solar mass
of \Ni\ from its decay to \Co\ and subsequent decay to \Fe :
$\dot{S} = 6.31 \times 10^{43} e^{-t_R/8.8} + 1.43 \times 10^{43}
e^{-t_R/111}\,\, \rm erg\,\, s^{-1}\,\, M_{\odot}^{-1}$,
where $t_R$ is the time in days for the supernova to rise from
explosion to maximum light.  Using $t_R=s \times 19.5$
days\citep{2006conley}, for SNLS-03D3bb, $t_R=22$ days (see
SI for the effect of a shorter rise).  We use
$\alpha =1.2$ as a conservative value, although for high \Ni\ masses,
$\alpha$ may be lower, since nickel above the photosphere will not
contribute to the luminosity\citep{1995PhRvL..75..394N}.

\begin{figure*}
\includegraphics[width=6in]{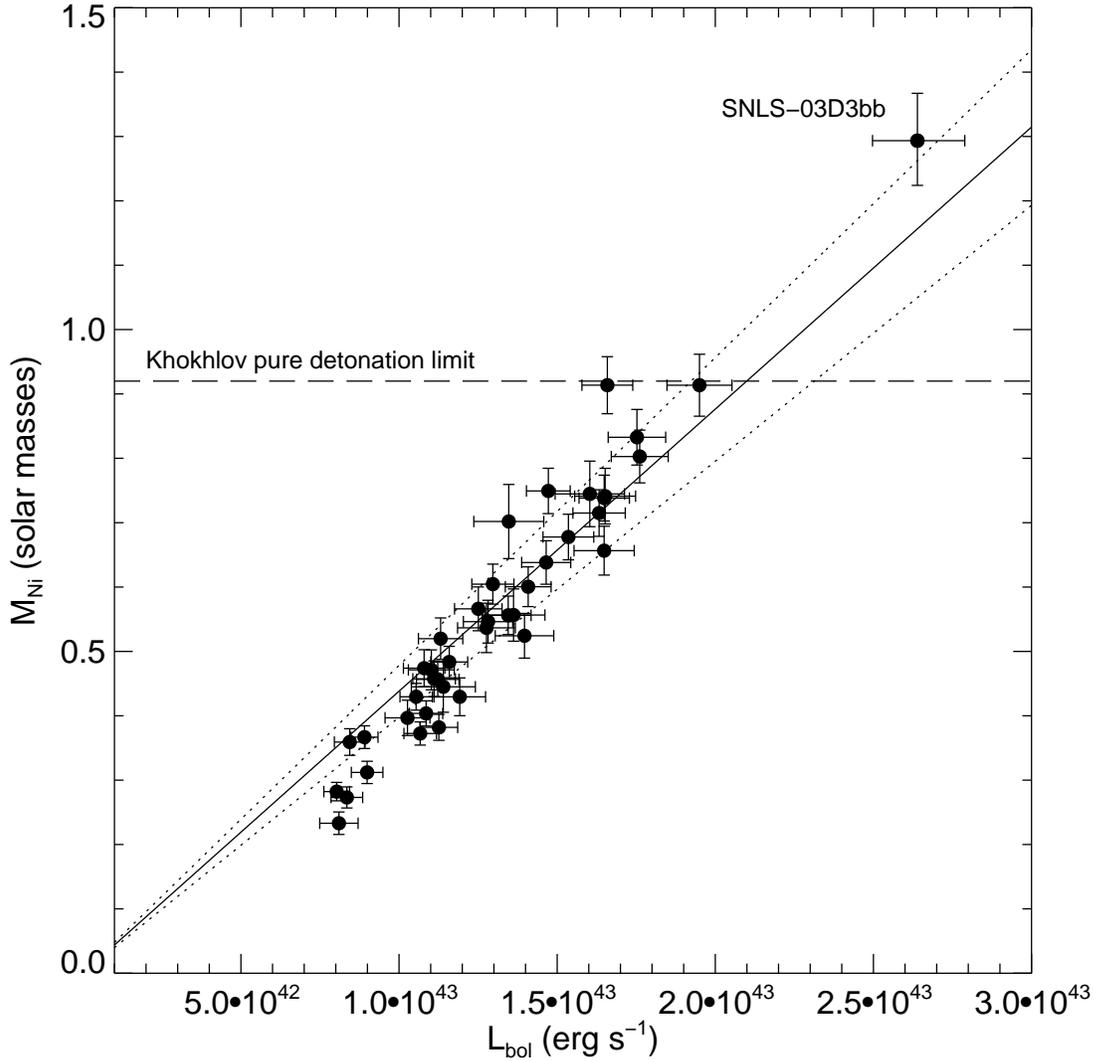}
\caption{Bolometric luminosity vs. implied \Ni\ mass for SNLS-03D3bb
  and low redshift SNe Ia\citep{2006A&A...447...31A}.  The low redshift
  SNe Ia were fit using the same techniques as those used for
  SNLS-03D3bb: the bolometric luminosity was determined using the peak
  magnitude in the $V$ band from a simultaneous fit to $B$ and $V$
  band data.  For the low redshift SNe Ia, we integrated the s=1 SN
  template\citep{2002PASP..114..803N} to obtain a bolometric
  correction of 0.06 magnitudes and an uncertainty (s.d.) of 0.05 mag
  for the combined bolometric and k-correction error.
  The solid line represents a normal stretch=1 SN Ia, with a rise time
  ($t_r$) of -19.5 days, while dotted lines show $s=0.9$ ($t_r=17.6$)
  and $s=1.1$ ($t_r=21.5$).  Low luminosity SNe Ia have lower
  stretches, and thus shorter rise times, resulting in less \Ni\ for a
  given luminosity, while high stretch SNe Ia show opposite behaviour.
  The dashed line shows an upper limit for the expected \Ni\ mass in a
  Chandrasekhar mass SN~Ia, obtained by burning the entire white
  dwarf to iron-peak elements in a
  detonation\citep{1993A&A...270..223K}.
\label{arnett}} 
\end{figure*}

\begin{figure*}
\includegraphics[width=6in]{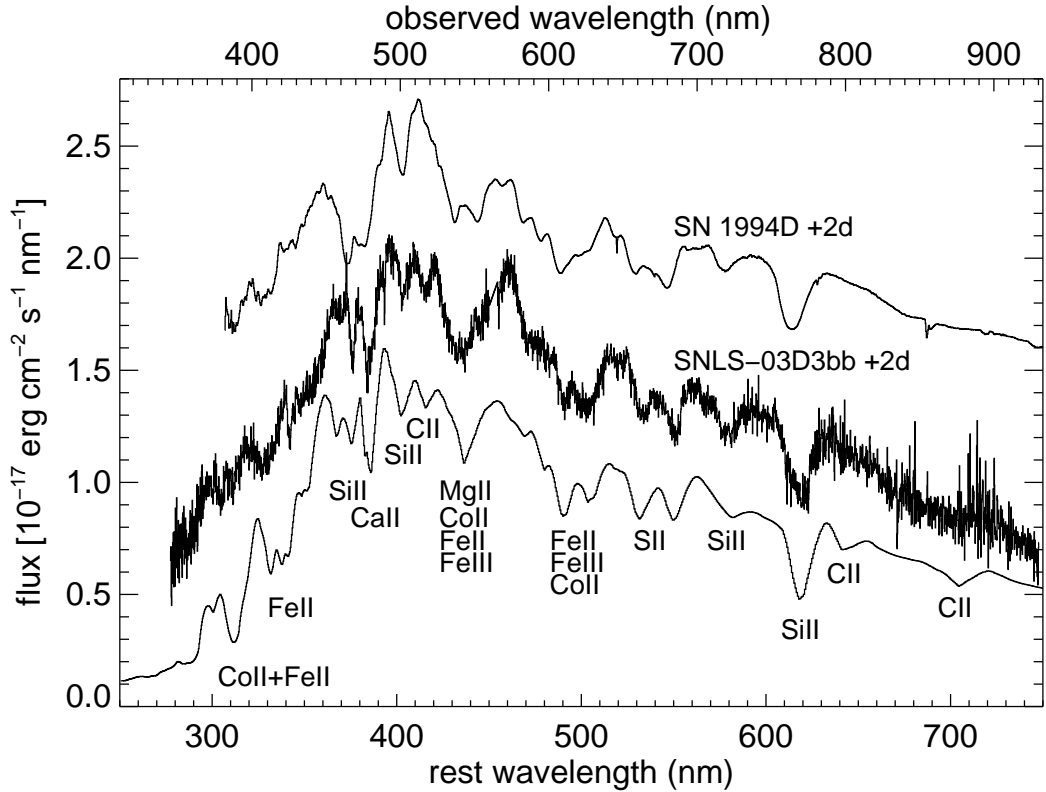}
\caption{Keck LRIS spectrum of SNLS-03D3bb at 2 days after
  maximum light compared to a spectrum of the normal Type Ia SN~1994D.
  Also plotted is a SYNOW fit to the data with dominant ions labeled.
  SYNOW is a parameterized resonance-scattering code, allowing the
  user to adjust optical depths, temperatures, and velocities to aid
  in the identification of supernova lines
  \citep{1997ApJ...481L..89F}.  SYNOW parameters are listed in the
  supplementary information.  SNLS-03D3bb shows the lines of IMEs
  typically seen in a SN Ia at maximum light --- SiII, SII, and CaII,
  but in SNLS-03D3bb the velocity of the lines is lower than usual.
  The line at 415 nm appears to be CII, but the other predicted
  carbon features cannot be clearly identified due to the lower
  signal-to-noise ratio of the spectrum in the red.  No other
  identification could be found for the 415 nm feature.
\label{bbsyn}}
\end{figure*}

To convert our $V$ magnitude into a bolometric equivalent, we use a
synthetic spectrum calculated to match the observed UV+optical
spectrum (Fig.~\ref{bbsyn}), but extended into the infrared (13\% of
the bolometric luminosity is from the IR extrapolation).  The
bolometric correction ($m_{\rm bc}$) is $0.07 \pm 0.03$ mag, such that
$M_{\rm bol} = M_V + m_{\rm bc}=-19.87\pm0.06$ mag.  Using these
numbers, we calculate $M_{\rm Ni} = 1.29\pm0.07$ \Msun\ for
SNLS-03D3bb, in agreement with the simple scaling argument used
earlier.  The quoted error is from the statistical,
k-correction, and bolometric correction errors added in quadrature.
SNLS-03D3bb has a significantly larger bolometric luminosity and 
implied \Ni\ mass compared to low redshift SNe (Fig.~\ref{arnett}).

\begin{figure*}
\includegraphics[width=6in]{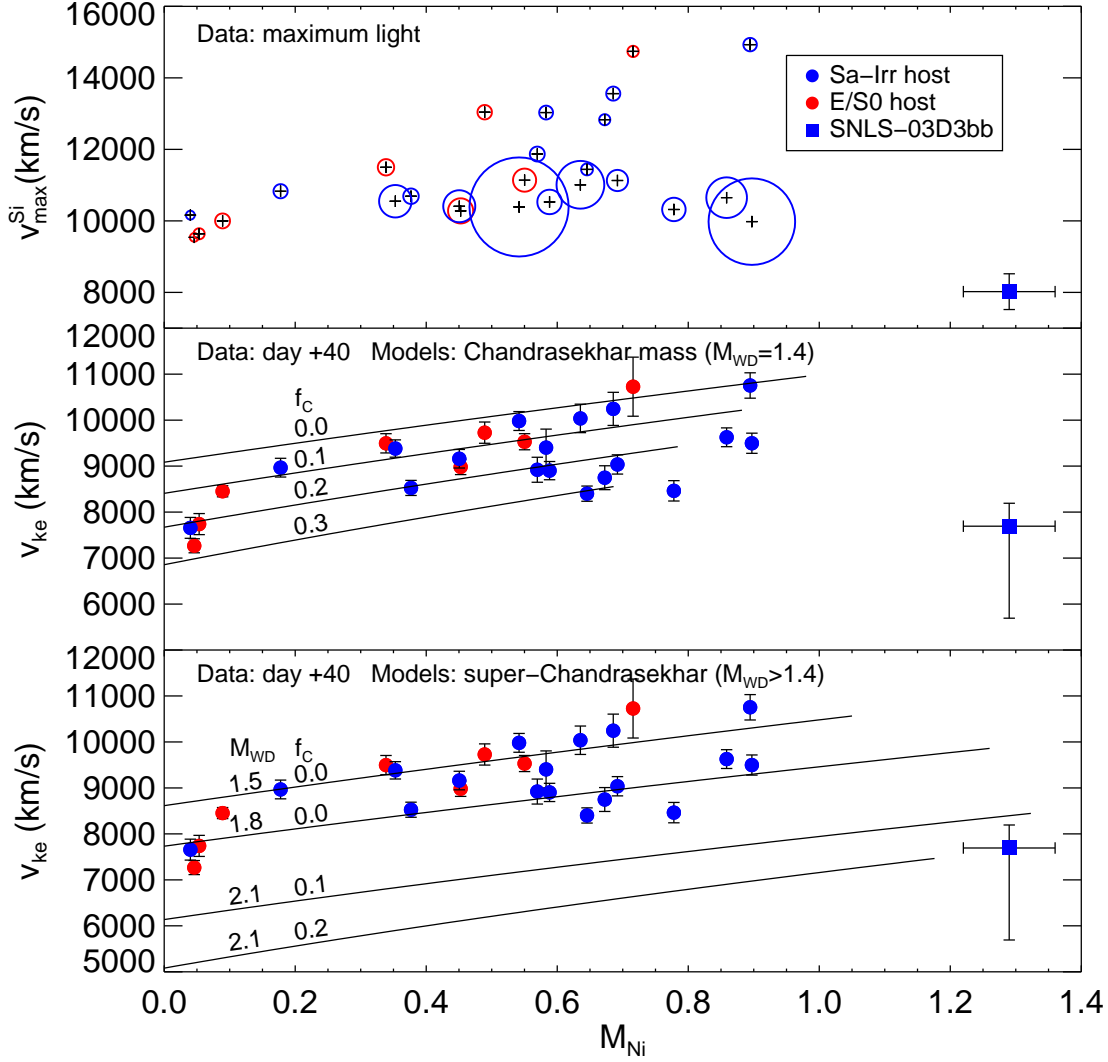}
\caption{Inferred Ni mass vs. SiII 615 nm velocity. {\bf a}, The data at maximum
  light\citep{2005ApJ...623.1011B}.  Ni masses are
  converted as described in the text using $M_{\rm bol}=M_{B}+0.2$.
  Red circles are from early type galaxies (E or S0), while blue
  circles are from late type galaxies (Sa-Irr).  Circle size is
  proportional to $v_{\rm Si}/\dot{v}_{\rm Si}$, where $\dot{v}_{\rm
    Si}$ is the rate of change of the velocity of the SiII feature.  There is no
  $\dot{v}_{\rm Si}$ measurement for SNLS-03D3bb.  {\bf b},
  Kinetic energy velocity of SNe Ia versus Ni mass for 1.4 solar mass
  models with different fractions of unburned carbon ($f_C$).  This
  unburned fraction should not be much higher than $\sim 20\%$ because
  carbon is rarely seen in SN Ia spectra\citep{2006astro.ph..1614M}.
  Overplotted symbols are $v_{\rm Si}$ for low redshift 
  SNe Ia\citep{2005ApJ...623.1011B} extrapolated to 40 days after maximum (correcting for stretch).
  For SNLS-03D3bb we use $\dot{v}_{\rm Si}$ from its closest neighbor.
  The error bar reflects the range if an average value of
  $\dot{v}_{\rm Si}$ is used.  SNLS-03D3bb is not consistent with the
  1.4 \Msun\ model.  {\bf c}, As above, but showing that $M_{\rm
    WD} \sim 2$ \Msun\ models can explain SNLS-03D3bb.  Less extreme
  super-Chandrasekhar mass models are consistent with the low redshift
  data.  The three low \Ni\ SNe are not necessarily
  super-Chandrasekhar SNe Ia --- their large values of $\dot{v}_{\rm
    Si}$ make projections to 40 days uncertain.
\label{ben2p}
}
\end{figure*}

SNLS-03D3bb also has an unusually low ejecta velocity, as shown in the Keck
spetrum taken 2 days after maximum light (Fig. \ref{bbsyn}). 
With a SiII velocity of $8000 \pm 500$ \kms , it falls well outside the range of
velocities seen for this feature at maximum light 
(Fig. ~\ref{ben2p}).  This is hard to understand in the
Chandrasekhar mass model, which predicts {\em higher}
velocities for more luminous SNe Ia, in contrast to the unusually {\em
  low} velocities in SNLS-03D3bb. 

The kinetic energy ($E_k$) of a SN Ia arises from the difference
between the nuclear energy ($E_n$) obtained from the synthesis of
elements via fusion in the explosion and the binding energy ($E_b$) of
the white dwarf\citep{1992ApJ...392...35B}.  Thus the kinetic energy
velocity is: $v_{\rm ke}=\sqrt{\frac{2 (E_n-E_b)}{M_{\rm WD}}}$, where
$M_{\rm WD}$ is the mass of the white dwarf.  The binding energy of a
1.4 \Msun\ C/O WD is $0.5 \times 10^{51}$
ergs\citep{1984ApJ...286..644N}.  For a 2\Msun\ WD and a central
density of $4 \times 10^9$ g cm$^{-3}$, the binding energy is $1.3
\times 10^{51}$ ergs\citep{2005A&A...435..967Y}.

Since there are only three classes of elements in a SN Ia (iron-peak
elements, intermediate mass elements (IME), and unburned carbon and oxygen),
a simple model can be developed for the nuclear energy
generation, $E_n$.  Burning a mixture of equal parts carbon and oxygen
to the iron peak produces $E_{\rm Fe}=1.55 \times 10^{51}$ erg \Msun $^{-1}$,
while the synthesis of $^{28}$Si produces 76\% as much energy
\citep{1992ApJ...392...35B}.  Thus:
$E_n=E_{\rm Fe} M_{\rm WD} (f_{\rm Fe}+0.76 f_{\rm IME})$,
where $M_{\rm WD}$ is in solar masses, and $f_{\rm Fe}$ and $f_{\rm
  IME}$ are the fractional compositions of iron peak and intermediate
mass elements.  If $f_{\rm Fe}$ and $f_{\rm IME}$ do not sum to one,
the remainder is the fraction of unburned carbon and oxygen ($f_C$),
which does not contribute to the nuclear energy.  The \Ni\ makes up
approximately 70\% of iron-peak elements\citep{1984ApJ...286..644N,
1993A&A...270..223K}, so we adopt $M_{\rm Ni}=0.7 M_{\rm WD} f_{\rm
Fe}$, where $M_{\rm Ni}$ is the mass of \Ni .

In the Chandrasekhar mass model, more luminous SNe, with more
\Ni , have a higher $v_{\rm ke}$ (Fig.~\ref{ben2p}).
Increasing the fraction of unburned carbon and oxygen, $f_C$, can
lower the kinetic energy, perhaps accounting for some of the
dispersion in SN Ia velocities.  However, this also lowers the
available \Ni , so it cannot account for the low velocity 
seen in SNLS-03D3bb.

The kinetic energy gives the velocity of the supernova averaged over
the entire mass, approximately equivalent to the velocity a few weeks after
maximum light\citep{1992ApJ...392...35B}.  The most appropriate
observational signature of this velocity is unclear, since SN Ia line
velocities change with time, and different ions can have different
relative velocities.  We find good agreement between the SiII
velocity at 40 days after maximum light\citep{2005ApJ...623.1011B} and
the theoretical kinetic energy velocity, but we emphasize that this is
in imperfect comparison.

A super-Chandrasekhar mass reproduces the low velocities seen in 
SNLS-03D3bb (Fig.~\ref{ben2p}).  Since Chandrasekhar models with more Ni produce
higher velocities, the low velocities of SNLS-03D3bb imply an
increased progenitor binding energy and thus a
larger total mass.  
As a caveat, we note that this simple calculation is only intended to
illustrate general trends.  Future theoretical studies will have to
assess such complications as using different ions, different white
dwarf density structures, and a wider range of binding energies.

Super-Chandrasekhar mass SNe Ia should be more likely in a young
stellar population, where the most massive stars
exist\citep{1994MNRAS.268..871T,2001ApJ...554L.193H}.  The low mass,
star forming host of SNLS-03D3bb is consistent with this scenario (see SI).
Thus, the apparent existence of super-Chandrasekhar mass SNe Ia may
explain why the most luminous SNe Ia only occur in
young stellar environments\citep{1996AJ....112.2391H,2006sullivan}.
The standard Chandrasekhar mass model offers no explanation for this
behaviour, since the total amount of fuel and triggering mechanism
should be independent of the mass of the progenitor stars.

SNe such as SNLS-03D3bb will have to be screened out in cosmological
studies.  Since younger stellar environments produce more luminous
SNe, as the mean stellar age decreases with redshift the mean
properties of SNe Ia will change\citep{2006sullivan}.  This can be
calibrated if all SNe obey the same stretch-luminosity relationship,
but SNLS-03D3bb does not.  Its peculiarity was so obvious that it was
excluded from the SNLS cosmological result\citep{2006A&A...447...31A},
but less extreme objects could lurk in
SN samples.  Future cosmology studies will have to carefully
scrutinise SNe Ia from young populations to see if they obey the same
lightcurve shape-luminosity relationship as other SNe Ia.

\bibliography{astro}

{\bf Supplementary Information} is linked to the online version of
this paper at www.nature.com/nature.

{\bf Acknowledgements} SNLS relies on observations with MegaCam, a
joint project of CFHT and CEA/DAPNIA, at the Canada-France-Hawaii
Telescope (CFHT). We used data products produced at the Canadian
Astronomy Data Centre as part of the CFHT Legacy Survey.  Some data were
obtained at the W.M. Keck Observatory.  We acknowledge support 
from NSERC, NERSC, CIAR, CNRS/IN2P3, CNRS/INSU, CEA, and the DOE.  

{\bf Author Infomation} Reprints and permissions information is
available at\\ npg.nature.com/reprintsandpermissions.  The authors
declare no competing financial interests. Correspondence and requests
for materials should be addressed to D.A.H (howell@astro.utoronto.ca).

\newpage

\newpage

\newpage

\newpage

\end{document}